\documentclass{ifacconf}

\usepackage{natbib}        

\usepackage{xcolor,calc}
\usepackage[pdftex]{graphicx}
\usepackage{amssymb,amsmath,url}
\usepackage{algorithm}
\usepackage{bm} 
\usepackage{multirow}
\usepackage{booktabs}	
\usepackage{epstopdf}
\usepackage{color}
\usepackage{subfigure}
\usepackage{bigfoot}		
\usepackage{epstopdf}
\usepackage{listings}
\usepackage{comment}
\usepackage{setspace}

\theoremstyle{plain}
\newtheorem{theorem2}{Theorem}
\newtheorem{lemma}{Lemma}

\theoremstyle{definition}
\newtheorem{definition}{Definition}

\newtheorem{proof}{Proof}
\renewenvironment{proof}{{\bfseries Proof.}}{}

\newcommand{\R}{\mathbb{R}}

\newcommand{\X}{\mathbb{X}}

\newcommand*{\QEDA}{\hfill\ensuremath{\blacksquare}}

\DeclareMathOperator*{\argmin}{arg\,min}
\usepackage{bm}	
\usepackage{algorithm}
\usepackage[]{algpseudocode}

\allowdisplaybreaks
\begin{document}
\begin{frontmatter}

\title{No-Regret Learning from Partially Observed Data in Repeated Auctions\thanksref{footnoteinfo}} 

\thanks[footnoteinfo]{This research was gratefully funded by the European Union ERC Starting Grant CONENE, and the Swiss National Science
Foundation grant SNSF $200021\_172781$.}
\thanks[footnoteinfo2]{These authors contributed equally to this work.}

\author[First]{Orcun Karaca\thanksref{footnoteinfo2}} 
\author[First]{Pier Giuseppe Sessa\thanksref{footnoteinfo2}}
\author[First]{Anna Leidi} and
\author[First]{Maryam Kamgarpour}

\address[First]{Automatic Control Laboratory, ETH Z\"urich, Switzerland.\\
(e-mails: {{\{okaraca, sessap, leidia, maryamk\}@ethz.ch}}).}

\begin{abstract} 
We study a general class of repeated auctions, such as the ones found in electricity markets, as multi-agent games between the bidders.
In such a repeated setting, bidders can adapt their strategies online based on the data observed in the previous auction rounds.  
Moreover, if no-regret algorithms are employed by the bidders to update their strategies, the game is known to converge to a coarse-correlated equilibrium, which generalizes the notion of Nash equilibrium to a probabilistic view of the auction state.
Well-studied no-regret algorithms depend on the feedback information available at every round, and can be mainly distinguished as bandit (or payoff-based), and full-information. However, the information structure found in auctions lies in between these two models, since participants can often obtain partial observations of their utilities under different strategies. 
To this end, we modify existing bandit algorithms to exploit such additional information. Specifically, we utilize the feedback information that bidders can obtain when their bids are not accepted, and build a more accurate estimator of the utility vector. This results in improved regret guarantees compared to standard bandit algorithms. Moreover, we propose a heuristic method for auction settings where the proposed algorithm is not directly applicable. 
Finally, we demonstrate our findings on case studies based on realistic electricity market models.

\end{abstract}


\end{frontmatter}
\vspace{-.6cm}
\section{Introduction}
\vspace{-.15cm}
Auctions are effective tools for allocating resources and determining their values among a set of participants. To this end, many complex auction mechanisms have been proposed to buy and sell different resources. Prominent examples include those in spectrum auctions~\citep{cramton2013spectrum,bichler2017handbook}, electricity markets~\citep{ahlstrom2015evolution,cramton2017electricity}, auctions for fish harvesting rights and other natural capitals~\citep{bichler2019designing,teytelboym2019natural}.
Moreover, a periodic \emph{recurrence} is intrinsic to several applications. For instance, in electricity markets, the same participants are generally involved in similar market transactions day after day~\citep{bose2019some,abbaspourtorbati2015swiss}.

Repeated auctions can be studied as multi-agent games among different players, or bidders in the auction, equipped with adaptive and sequential bidding algorithms. In such repeated games, the performance of a player is typically measured in terms of \emph{regret}, which is the utility loss incurred when compared to the best fixed action over a sequence of rounds. 
Moreover, if all players exhibit no-regret, that is, a diminishing regret as the number of rounds increase, the game is known to reach a so-called coarse-correlated equilibrium. This equilibrium concept generalizes Nash equilibrium to the case where all players are endowed with a probability distribution over the state of the game, see~\citep{cesa2006prediction} and~\citep[Ch.~13]{roughgarden2016twenty}. 

Several no-regret learning algorithms exist, and both the algorithms and their performance crucially depend on the feedback information available at every round of the game. In the \emph{full-information} setting, where the player observes its utility under every action~\citep{freund1997decision}, such algorithms attain an optimal $\mathcal{O}(\sqrt{T\log K})$ regret, with~$T$ being the horizon length, and~$K$ being the number of actions. When only \emph{bandit} feedback is available, that is, the player only observes its utility for the action picked~\citep{auer2002nonstochastic}, the optimal regret is $\mathcal{O}(\sqrt{KT\log K})$ with a significantly worse dependence on the number of actions~$K$.
However, the information structure found in auctions lies in between these two models; on the one hand, full-information feedback is unrealistic since it requires perfect knowledge of opponents' bids and market constraints, and on the other hand, bandit feedback is too restrictive since partial observations of utilities under different actions are often available owing to the auction rules and the additional information released. 
For instance, in auctions of electrical power marginal prices are announced after every round~\citep{NYISO1,ENTSOE1}. 
Motivated by these particularities of auctions, our goal is to extend existing bandit algorithms to account for such additional information.

Our contributions are as follows. We consider no-regret learning in a general class of repeated auctions. Employing standard results from the multi-armed bandit literature, we propose an algorithm that participants could use to update their bidding strategies based on the observed auction data. Specifically, our approach exploits the information available to the bidders when their bids are not accepted by the auctioneer. We show that the proposed algorithm enjoys a no-regret guarantee that can strictly improve upon bandit algorithms. We demonstrate this fact on a market with a simple supply-demand balance constraint and a marginal pricing payment mechanism. {Then, we consider a more general market setting for which the algorithm is not directly applicable since it requires parameters that are unknown in this setting, that is, revelation probabilities for additional information.} We propose a heuristic method for this case and demonstrate it outperforms existing bandit algorithms in our experiments.

{Let us contrast our work with the existing works on general learning algorithms with partial information. A partial monitoring framework was introduced in \citep{Piccolboni2001} and was extensively studied (e.g., in \citet[\S 6.4]{cesa2006prediction}) for online learning with feedback matrices. However, this setting is restricted to learning problems where the utilities are chosen from a finite set, e.g., $\{0,1\}$.
More recently, 
\citet{mannor2011bandits} introduced a novel framework in which the learning agent is equipped with a sequence of feedback graphs over the action space encoding which additional information is revealed by every action. 
In case the feedback graph is known before picking the action, \citet{lykouris2017small} obtained small loss regret guarantees that do not depend on the number of actions.
In case the feedback graph is revealed only after picking the action, the works of \citep{alon2015online,alon2017nonstochastic} derived regret bounds as a function of the feedback graphs' independence numbers (or sizes of their maximum acyclic subgraphs).
In contrast to these works, in our auction framework such graphs may not be known to the bidders at any point during the game. For instance, in the general market problem considered in Section~\ref{sec:gen_markt} revelation probabilities for the additional information originate from a hidden feedback graph, and we have to develop meaningful heuristics to compute these parameters. 
Finally, partial information in the context of multi-agent learning have been explored in \citep{sessa2019no}, assuming players can observe their opponents' actions, in addition to the standard bandit feedback. Differently, in this work we do not assume observation of opponents' bids. }

The rest of the paper is organized as follows. Section~\ref{sec:pre} lays out the preliminaries for the auction framework and existing no-regret algorithms. Our algorithm is proposed in Section~\ref{sec:algo}, followed by two applications.
Section~\ref{sec:num} presents the case studies based on optimal power flow and the Swiss reserve market. Section~\ref{sec:conc} concludes the~paper.

\section{Preliminaries}\label{sec:pre}

\subsection{Repeated Auction Framework}

We consider a general reverse (or procurement) auction problem.
The set of participants consists of the bidders $\ell\in\mathcal{N}=\{1,\ldots,|\mathcal{N}|\}$. Let there be $q\in\mathbb{N}$ types of goods/supplies in the auction. For instance, in an electricity market, these types could refer to active and reactive power injections at different nodes, and also control reserves. Goods of the same type from different bidders are interchangeable to the auctioneer. We assume each bidder has a private true cost function $c_\ell:\X_\ell\rightarrow\R_+,$ $\X_\ell\subseteq\R_+^q.$ We further assume that $0\in\X_\ell$ and $c_\ell(0)=0$.\footnote{This holds for many electricity markets that do not allow shot-down costs, see~\citep{karaca2018core} for a discussion.} Each bidder~$\ell$ has a finite strategy set $\mathcal{K}_\ell=\{1,\ldots,|\mathcal{K}_\ell|\}$ that consists of its true cost function and bid functions of the form  $b_\ell^k:\X_\ell^k\rightarrow\R_+,$ $\X_\ell^k\subseteq\R_+^q,$ where $k\in\mathcal{K}_\ell,$ $0\in\X_\ell^k\subseteq\R_+^q,$ and $b_\ell^k(0)=0.$ Without loss of generality, $1\in\mathcal{K}_\ell$ corresponds to the true cost function.

Let $T\in\mathbb{N}$ be the horizon length. We assume that for all rounds the true costs and the strategy set $\mathcal{K}_\ell$ remain unchanged. Let $k_\ell(t)\in\mathcal{K}_\ell$ denote the strategy of bidder~$\ell$ at time $t\leq T.$ Given the strategy profile $\mathcal{B}(t)=\{b_\ell^{k_\ell(t)}\}_{\ell\in\mathcal{N}}$, a \textit{mechanism} defines an \textit{allocation rule} $x_\ell^*(\mathcal{B}(t))\in\X_\ell^{k_\ell(t)},$ and a \textit{payment rule} $p_\ell(\mathcal{B}(t))\in\R$ for each bidder $\ell.$ In many auctions, the allocation rule is determined by an optimization problem of the form,
\begin{equation}\label{eq:market_prob}
    \begin{split}
        J(\mathcal{B}(t))= &\min_{x\in\X(t)} \sum_{\ell\in\mathcal{N}}b_\ell^{k_\ell(t)}(x_\ell)\\
        &\ \ \mathrm{s.t.}\ \sum_{\ell\in\mathcal{N}} x_\ell \in \mathbb{S}\, ,
    \end{split}
\end{equation}
where $\X(t)=\prod_{\ell\in\mathcal{N}}\X_\ell^{k_\ell(t)}\subset \mathbb{R}_+^{q|\mathcal{N}|}$, and the set $\mathbb{S}\subset \mathbb{R}^q_+$ corresponds to the market constraints. In an electricity market, these constraints may correspond to network balance constraints found in optimal power flow problems~\citep{wu1996folk}, or probabilistic security requirements found in control reserves markets~\citep{abbaspourtorbati2015swiss}.
With the definition above, we also assume that the market constraints remain unchanged throughout the horizon. 

Let the optimal solution of \eqref{eq:market_prob} be denoted by $x^*(\mathcal{B}(t))$. We assume that in case of multiple optima there is a tie-breaking rule. 
The utility of bidder~$\ell$ is linear in the payment received; $u_\ell(t)=u_\ell(\mathcal{B}(t))=p_\ell(\mathcal{B}(t))-c_\ell(x_\ell^*(\mathcal{B}(t))).$ A bidder whose bid is not accepted, $x_\ell^*(\mathcal{B}(t))=0$, is not paid, $u_\ell(t)=0$, and is referred to as \emph{loser}. 

The fundamental goal in auction theory is efficiency, which is attained when problem \eqref{eq:market_prob} is solved under the condition that the bidders submitted their true costs~\citep{krishna2009auction}. Since the bidders strategize to receive larger payments, there has been many proposals for different payment rules, for instance, the pay-as-bid~\citep{bernheim1986menu,karaca2017game}, locational marginal pricing~\citep{schweppe2013spot}, Vickrey-Clarke-Groves (VCG)~\citep{krishna2009auction,pgs2017}, and core-selecting mechanisms~\citep{day2008core}. Even though these payment rules are well-discussed in terms of how the bidders should pick their strategies in a Nash equilibrium, in reality bidders are profit maximizing entities that compete under privacy considerations and limited information. Hence, it may not be realistic to assume that they can compute or they would be willing to pick their Nash equilibrium strategy.\footnote{Under the VCG mechanism, reporting true cost is the dominant strategy of every bidder. However, bidders can still turn to group deviations to maximize their profits~\citep{ausubel2006lovely}.} Instead, a more practical assumption is that the bidders choose their strategies following simple adaptive and sequential algorithms based on observed auction data.
\vspace{-.1cm}
\subsection{No-Regret Learning and Correlated Equilibrium}
\vspace{-.1cm}
The performance of a generic bidder~$\ell$ in a repeated auction can be measured in terms of \emph{regret}, which is a standard notion used in the online learning literature to measure the performance of a sequential decision making algorithm (see, e.g., \citep{shalev2012online,cesa2006prediction}). For the following, define $\mathcal{B}_{-\ell}(t)=\{b_j^{k_j(t)}\}_{j\in\mathcal{N}\setminus\{\ell\}}.$
\begin{definition}[Regret]\label{def:reg} The \emph{regret} of bidder~$\ell$ at time $T$ is
\begin{equation*}
    R_\ell(T) = \max_{k\in \mathcal{K}_\ell}\,\sum_{t=1}^T u_\ell(\lbrace \mathcal{B}_{-\ell}(t), b_\ell^k\rbrace) - \sum_{t=1}^T u_\ell(t) \,.
\end{equation*}
\end{definition}

After $T$ rounds, hence, $R_\ell(T)$ quantifies the maximum profit bidder~$\ell$ could have made had she known the sequence of opponents' bids ahead of time, and had she chosen the best fixed strategy in $\mathcal{K}_\ell$. 
An algorithm for bidder~$\ell$ is \emph{no-regret} if $R_\ell(T)/T \rightarrow 0$ as $T \rightarrow \infty$.

In a multi-agent setting like the aforementioned auction problems, the notion of regret has a close connection to equilibria. In fact, it can be shown that if every participant bids according to a no-regret algorithm, the empirical distribution of bids converge to a \emph{coarse-correlated equilibrium} (CCE) of the one-shot game \citep{cesa2006prediction}.
\begin{definition}[CCE] A \emph{coarse-correlated equilibrium} is a distribution $\sigma$ over $\prod_{\ell\in\mathcal{N}}\mathcal{K}_\ell$ such that, for each bidder $l\in \mathcal{N}$,
\begin{equation*}
    \mathbb{E}_{\mathcal{B}\sim \sigma}\left[u_\ell(\mathcal{B})\right] \geq \mathbb{E}_{\mathcal{B}\sim \sigma}\left[u_\ell(\lbrace \mathcal{B}_{-\ell}, b_\ell^{k} \rbrace)\right], \quad \forall k \in \mathcal{K}_\ell \,.
\end{equation*}
\end{definition}

As a remark, CCEs are the largest (and the weakest) class of equilibria, and include pure and mixed Nash Equilibria (NE). However, computing any NE is PPAD-complete {(a weaker version of NP-completeness, see~\citep{Daskalakis2009})} and requires full knowledge of the game.\looseness=-1

No-regret algorithms for a generic bidder~$\ell$ can be derived by mapping the repeated auction to an online learning problem faced by bidder $\ell
$: At every round $t$ bidder~$\ell$ picks an action $k(t) \in  \mathcal{K}_\ell$, an adversarial environment selects a \emph{loss vector} $\mathbf{l}_t \in [0,\,1]^{|\mathcal{K}_\ell|}$, and bidder~$\ell$ incurs loss $\mathbf{l}_t[k(t)]$. 
Since the bidders are in general aware of the range of their utilities, we assume they can map their utilities to losses in the $[0,\,1]$ interval. Thus, we let utility $u_\ell(t)=u_\ell(\mathcal{B}(t))\in \R$ correspond to loss $l_\ell(t)=l_\ell(\mathcal{B}(t)) = 1 - s_\ell(u_\ell(\mathcal{B}(t))) \in [0,1]$, where $s_\ell:\R\rightarrow[0,\,1]$ is a suitable monotone map. At every round $t$, hence, the corresponding loss vector is:
\begin{equation}\label{eq:full_info_feedback}
\mathbf{l}_t = \left[l_\ell(\lbrace \mathcal{B}_{-\ell}(t), b_\ell^1\rbrace),\ldots, l_\ell(\lbrace \mathcal{B}_{-\ell}(t), b_\ell^{|\mathcal{K}_\ell|}\rbrace)\right]\,.
\end{equation}
Note that the monotonicity of $s_l$ implies that our regret definition maps to a regret based on such loss formulation.

 In order to attain no-regret, bidder~$\ell$ must randomize her actions and bid according to \emph{mixed} strategies, that is, probability distributions over $\mathcal{K}_\ell$ \citep{cesa2006prediction}. 
{Thus, one often reasons about regrets in expectation.}
Let $K= |\mathcal{K}_\ell|$ be the number of actions in the strategy set of bidder~$\ell$ (we drop the dependence on $\ell$ for ease of notation) and $\mathbf{w}_t \in [0,1]^K$ be the mixed strategy of bidder~$\ell$ at round $t$.
 A large family of no-regret algorithms are based on a simple, yet effective, Multiplicative Weight Update (MWU) rule to update bidder $l$'s mixed strategy \citep{freund1997decision}. Such update rule is summarized in Algorithm~\ref{alg:MWU}. At every time $t$, $\mathbf{w}_{t+1}$ is computed proportionally to $\mathbf{w}_{t}$ via an estimate $\tilde{\mathbf{l}}_t$ of the loss vector $\mathbf{l}_t$. The performance of such algorithms depend on the chosen estimate. 

\begin{algorithm}[t!]
    \caption{MWU algorithm for bidder $l$} \label{alg:MWU}
\begin{algorithmic}
\State \textbf{Input:} Strategy set $\mathcal{K}_\ell$ with $|\mathcal{K}_\ell|=K$, parameter $\eta$
\State Initialize mixed strategy $\mathbf{w}_1 = [^1/_{K}, \ldots,^1/_{K}]$
\For {$t = 1,2,\dotsc, T$}
\State Compute estimate $\tilde{\mathbf{l}}_t$ of the loss vector $\mathbf{l}_t \in [0,1]^{K}$
\State Update mixed strategy:
    $$\mathbf{w}_{t+1}[i] \propto \mathbf{w}_{t}[i] \cdot \exp \big(- \eta \: \tilde{\mathbf{l}}_t[i]  \big), \quad i=1,\ldots, K$$
\EndFor
\end{algorithmic}
\end{algorithm}

Full-information feedback algorithms such as \textsc{Hedge}~\citep{freund1997decision} use the true loss vector $\mathbf{l}_t$ (which in our case corresponds to \eqref{eq:full_info_feedback}) as the estimate $\tilde{\mathbf{l}}_t$.
Such algorithms attain an optimal expected regret of $\mathcal{O}(\sqrt{T\log K})$.
However, full-information feedback is unrealistic in repeated auctions, since computing $l_\ell(\lbrace \mathcal{B}_{-\ell}(t), b_\ell^k\rbrace)$ for $k \neq k_\ell(t)$ requires the full knowledge of the bids $\mathcal{B}_{-\ell}(t)$ and the market constraints to generate solutions of the optimization problem~\eqref{eq:market_prob}.

Bandit algorithms such as \textsc{Exp3}~\citep{auer2002nonstochastic} use only the obtained loss $l_\ell(t)$ to build an estimate of the loss vector as:
\begin{equation}\label{eq:exp3_estimator}
\tilde{\mathbf{l}}_t = \left[0,\ldots, 0 , \frac{l_\ell(t)}{\mathbf{w}_t[k_\ell(t)]} ,0,\ldots, 0 \right] \,.
\end{equation}
Although one can show that $\tilde{\mathbf{l}}_t$ is an \emph{unbiased} estimate of the true loss vector $\mathbf{l}_t$, that is, $\mathbb{E}_{k \sim \mathbf{w}_t}\big[\tilde{\mathbf{l}}_t[i] \big] = \mathbf{l}_t[i]$ for all $i\in \mathcal{K}_\ell$, the variance of $\tilde{\mathbf{l}}_t[i]$ grows with the inverse of the squared probability $\mathbf{w}_t[i]^2$. 
This leads to an expected regret of $\mathcal{O}(\sqrt{K T\log K})$ (matching an algorithm-independent lower bound \citep{auer2002nonstochastic} up to log factors), which scales significantly worse with the number of actions $K$ and often leads to poor performance.

\vspace{-.1cm}
\section{No-Regret Learning from Partially Observed Data}\label{sec:algo}
\vspace{-.1cm}
While the full-information feedback is unrealistic, the feedback information available in repeated auctions is typically richer than the bandit feedback. In fact, when bidder~$\ell$ is a \emph{loser} of the auction, not only she observes the loss for the chosen bid, which corresponds to $l_\ell(0)$, but often can also infer about other actions that would have led to such losing outcome. This is the key idea that we exploit to improve upon the regret bound of the existing bandit \textsc{Exp3} algorithm. We will then make it more concrete in Sections~\ref{sec:simp_markt} and~\ref{sec:gen_markt} by considering two specific classes of auctions.
As a remark, when bidder $\ell$ is a winner of the auction she can often infer about other winning actions; however, she cannot infer the exact utilities. We plan on addressing this fact in our future work.

\vspace{-.1cm}
\subsection{Extended \textsc{Exp3} Algorithm}
\vspace{-.1cm}
Consider time $t$, and assume bidder~$\ell$ loses the auction, that is, $x^*(\mathcal{B}(t))=0$. In this case, we assume bidder~$\ell$ gets to know of a subset $\mathcal{L}_t\subseteq\mathcal{K}_\ell$ of losing actions (including $k_\ell(t))$), i.e., actions which would have led to the same losing outcome. 
In the class of auctions considered in Sections~\ref{sec:simp_markt}, for instance, bidder~$\ell$ finds out about $\mathcal{L}_t$ from the marginal price set by the auctioneer.

For any action $k \in \mathcal{K}_\ell$, we let $\mathbf{r}_t[k]$ denote the probability that bidder~$\ell$ discovers the utility of action~$k$. We refer to $\mathbf{r}_t[k]$'s as the \emph{revelation probabilities}. Note that $\mathbf{r}_t[k]$ is always greater than the probability $\mathbf{w}_t[k]$ of playing action~$k$, since when action $k$ is played, bidder~$\ell$ directly discovers its utility. However, as explained before, bidder~$\ell$ can find out about losing actions even when playing actions $i\neq k$. We will provide an explicit expression of such probabilities for the auctions considered in Sections~\ref{sec:simp_markt} and \ref{sec:gen_markt}. 
For now, suppose such revelation probabilities are known to bidder~$\ell$. 
Using this information, bidder~$\ell$ can construct the following estimator, for any $k\in\mathcal{K}_\ell$:
\begin{equation}\label{eq:our_estimator}
 \tilde{\mathbf{l}}_t[k] = \begin{cases}\frac{l_\ell(t)}{\mathbf{r}_t[k]},&\text{if }x_\ell^*(\mathcal{B}(t))>0 \text{ and } k=k_\ell(t), \\
 0,&\text{if } x_\ell^*(\mathcal{B}(t))>0 \text{ and } k\neq k_\ell(t), \\
 \frac{l_\ell(0)}{\mathbf{r}_t[k]},&\text{if } x_\ell^*(\mathcal{B}(t)) = 0  \text{ and } k \in \mathcal{L}_t,\\
 0,&\text{if } x_\ell^*(\mathcal{B}(t))=0 \text{ and } k \notin \mathcal{L}_t \,,
 \end{cases}
\end{equation}
where $l_\ell(0)=1 - s_\ell(0)$ for the sake of brevity, and $\mathbf{r}_t[k]=\mathbf{w}_t[k]$ for the first case in which the bidder is winning. 
In words, if action $k_\ell(t)$ is a winning action, the estimator~\eqref{eq:our_estimator} coincides with the bandit feedback estimator~\eqref{eq:exp3_estimator}. On the other hand, if $k_\ell(t)$ is a losing action, $\tilde{\mathbf{l}}_t$ computed above has a larger number of non-zero entries compared to \eqref{eq:exp3_estimator}. 

It is not hard to prove the following fact, which is the key to obtaining a sublinear regret bound. 

\begin{fact}\label{fact}
The estimator $\tilde{\mathbf{l}}_t$ computed as in \eqref{eq:our_estimator} is an unbiased estimate of the true loss vector $\mathbf{l}_t$ in \eqref{eq:full_info_feedback}.
\end{fact}
\vspace{-.1cm}

\begin{proof}
We prove that, at any given time $t$, $\mathbb{E}_{k\sim \mathbf{w}_t} \big[ \tilde{\mathbf{l}}_t[i] \big] = \mathbf{l}_t[i]$ for every action $i \in \mathcal{K}_\ell$.
First, consider any $i$ such that $x^*(\lbrace \mathcal{B}_{-\ell}(t), b_\ell^i\rbrace) > 0$ (winning action). We have: 
\begin{align*} \mathbb{E}_{k\sim \mathbf{w}_t} \big[ \tilde{\mathbf{l}}_t[i ] \big] 
=   \mathbf{w}_t[i] \cdot \frac{l_\ell(t)}{\mathbf{w}_t[i]} + (1-\mathbf{w}_t[i]) \cdot 0  = \mathbf{l}_t[i] \,,
\end{align*}
since loss $\mathbf{l}_t[i]$ is revealed only when action $i$ is sampled, hence when $l_\ell(t) = \mathbf{l}_t[i]$.
Then, consider actions $i$ such that $x^*(\lbrace \mathcal{B}_{-\ell}(t), b_\ell^i\rbrace) = 0$ (losing actions). We have:
\begin{align*} \mathbb{E}_{k\sim \mathbf{w}_t} \big[ \tilde{\mathbf{l}}_t[i ] \big] 
=   \mathbf{r}_t[i] \cdot \frac{l_\ell(0)}{\mathbf{r}_t[i]} + (1-\mathbf{r}_t[i]) \cdot 0  = \mathbf{l}_t[i] \,,
\end{align*}
since $\mathbf{r}_t[i]$ is by definition the probability that $i$ is revealed to be a losing action. Moreover, $ \mathbf{l}_t[i] = l_\ell(0)$ follows from $i$ being a losing action.
\QEDA
\end{proof}

Compared to the standard bandit estimator in~\eqref{eq:exp3_estimator}, the estimator considered above has a strictly smaller variance (since $\mathbf{r}_t[k] \geq \mathbf{w}_t[k]$ for all $k$) by virtue of the additional information used. 
To quantify such improvement, we define the quantity
$\alpha_t^k := \frac{\mathbf{r}_t[k]}{\mathbf{w}_t[k]}.$
It measures the additional knowledge available to bidder~$\ell$ about the loss $\mathbf{l}_t[k]$, when compared to the standard bandit feedback. Note that $\alpha_t^k \geq 1$, 
and it increases with the information available to bidder $l$.
Moreover, to make our results explicit, we define the following aggregate quantity, called the \emph{average feedback information}.

\begin{definition}[Average feedback information]\label{def:avg_add_info}
The~average feedback information available to bidder~$\ell$ over $T$ auction rounds is
$$ \alpha_\text{avg} := \left(\frac{1}{TK}\sum_{t=1}^T \sum_{k=1}^K \frac{1}{\alpha_t^k} \right)^{-1} \ \in [1,\,K] \,.$$
 \end{definition}
 
Bandit feedback corresponds to $\alpha_\text{avg}= 1$, since $\mathbf{r}_t[k] = \mathbf{w}_t[k]$ for all $k$, while $\alpha_\text{avg}= K$ in case of full-information feedback, since $\mathbf{r}_t[k] = 1 $ for all $k$.
{In repeated auctions, (based on the previous discussions) the average feedback information $\alpha_\text{avg}$ depends on the auction type and also the information available to the participants, and can be computed only after all the auction rounds are completed.}

The following theorem bounds the performance of the proposed algorithm, as a function of $\alpha_\text{avg}$.
\begin{theorem2}\label{thm:main}
Assume bidder~$\ell$ bids according to Algorithm~\ref{alg:MWU} with loss vector estimate $\tilde{\mathbf{l}}_t$ computed as in \eqref{eq:our_estimator} and $\eta = \sqrt{2 \alpha_\text{avg} \log(K) /(K T)}$, then 
$$ \mathbb{E} \big[ R_\ell(T) \big]\leq \sqrt{2\;(K/\alpha_\text{avg}) T \log K} \,. $$
\end{theorem2}

Theorem~\ref{thm:main} generalizes the regret bounds of full-information and bandit algorithms, obtained by setting $\alpha_\text{avg}= K$ or $\alpha_\text{avg}= 1$ respectively, to the case where intermediate information is available to bidder~$\ell$. In a repeated auction, where $\alpha_\text{avg}$ is typically greater than $1$, the obtained regret bound strictly improves  upon standard bandit guarantees.
Our proof is relegated to Appendix~\ref{app:A} and it follows from standard online learning arguments by making use of Fact~\ref{fact} and Definition~\ref{def:avg_add_info}.
To optimally select the learning rate $\eta$, the value $\alpha_\text{avg}$ needs to be known ahead of time. However, we expect that the bidders could estimate this value from the previous auction rounds.
In practice, in our experiments we show that a wide-range of $\eta$ leads to desirable performance.

{In comparison with the works considering feedback graphs \citep{alon2015online,alon2017nonstochastic,lykouris2017small}, {Theorem~\ref{thm:main} does not depend on any graph-theoretic quantity; instead, it depends on the notion of average feedback information defined above. This makes it applicable to our more general auction framework in that the way the additional information originates can be unknown to the bidders} (see Section~\ref{sec:gen_markt}).
Comparing with their algorithms and regret bounds, \citet{alon2017nonstochastic} use an estimator that coincides with \eqref{eq:our_estimator} where the revelation probabilities are computed from the feedback graph, and obtain the regret bound $\sqrt{2\sum_{t=1}^T m_t\log K}$, where $m_t$ is a precomputed upper bound on the maximum size of acyclic subgraphs of the feedback graph at round~$t$. On the other hand, \citep{lykouris2017small} requires knowing the feedback graphs ahead of time and relies on a different algorithm which is a freezing modification to \textsc{Hedge}. Their regret bound holds with high probability and it is of the form $o(\beta L^*),$ where $L^*$ is the loss of the best fixed strategy, and $\beta$ is an upper bound on the independence number of the feedback graphs at all rounds. Finally, note that these works replace the $\sqrt{K}$ term of the regret bounds of bandit algorithms with a graph-theoretic quantity, which might still be equal to $\sqrt{K}$ in the worst-case.}
\looseness=-1

\vspace{-.1cm}
\subsection{Auctions with Simple Constraints and Convex Bids}\label{sec:simp_markt}
\vspace{-.1cm}
We consider a simpler auction problem where the auctioneer has to procure a fixed amount $Q\in\R_+$ of a single type of good. In addition to the assumptions in Section~\ref{sec:pre}, each bidder is now equipped with a finite strategy set consisting of strongly convex and increasing cost/bid functions\footnote{Here, strong convexity is assumed to eliminate ties. Our results can easily be generalized to the convex setting. } over compact intervals. 
The allocation rule is given by,
\begin{subequations}\label{eq:simple_markt}
   \begin{align}
        x^*(\mathcal{B}(t))=&\argmin_{x\in\X(t)} \sum_{\ell\in\mathcal{N}}b_\ell^{k_\ell(t)}(x_\ell)\\
        &\quad \mathrm{s.t.}\quad \sum_{\ell\in\mathcal{N}} x_\ell \geq Q\, .\label{eq:simple_const}
    \end{align}
\end{subequations}
As a remark, some of the European reserve and day-ahead markets belong to this class since they ignore the network constraints.
Let $\lambda^*(\mathcal{B}(t))\in\R_+$ denote the Lagrange multiplier associated with the constraint in \eqref{eq:simple_const}, called the marginal price, and it is announced after each round~$t$ to all bidders. The payment rule is then given by
$p_\ell(\mathcal{B}(t))= \lambda^*(\mathcal{B}(t))\,x_\ell^*(\mathcal{B}(t)).$

Next, assume in round~$t$ bidder~$\ell$ is a loser with $x_\ell^*(\mathcal{B}(t))=0$. From the Karush-Kuhn-Tucker optimality conditions and strong duality of  \eqref{eq:simple_markt}~\citep{bertsekas1999nonlinear}, for any $k\in\mathcal{K}_\ell,$ we have $$\dfrac{d}{dx}{b_\ell^k}(x)\Big\rvert_{x=0}\geq\lambda^*(\mathcal{B}(t))\iff x_\ell^*(\lbrace \mathcal{B}_{-l}(t), b_\ell^{k} \rbrace)=0\,.$$
Equivalently, the marginal price information enables a losing bidder to be aware of all the other losing actions in its strategy set.\footnote{Similar arguments can be made for any Lagrange multiplier-based payments under strong duality and strong convexity.} Hence, bidder~$\ell$ gets to known the set of all losing actions: $$\mathcal{L}_t=\{k\in\mathcal{K}_\ell\,\rvert\,\dfrac{d}{dx}{b_\ell^k}(x)\Big\rvert_{x=0}\geq\lambda^*(\mathcal{B}(t))\}\,.$$
Notice that the set above is meaningful only when bidder~$\ell$ is a loser, since otherwise marginal price can potentially change whenever bidder~$\ell$ picks another action.

In the partially observed information setting
described above, a generic losing action $k$ is revealed when any other losing action is played, since each losing action reveals the full set of losing actions $\mathcal{L}_t.$ Hence, the revelation probabilities are given by:
\begin{equation*}
 \mathbf{r}_t[k] = \begin{cases}\sum_{j\in\mathcal{L}_t} \mathbf{w}_t[j],&\text{if }x_\ell^*(\mathcal{B}(t))=0\text{ and } k\in\mathcal{L}_t,\\
 \mathbf{w}_t[k],&\text{otherwise.}
 \end{cases}
\end{equation*}
{Observe that, in the first case, the probabilities are calculated by a sum, since choosing different losing actions constitute independent events.}
Then, the estimator $\tilde{\mathbf{l}}_t$ in~\eqref{eq:our_estimator} can be obtained by plugging the probabilities defined~above.
\looseness=-1

\vspace{-.2cm}
\subsection{General Auctions}\label{sec:gen_markt}
\vspace{-.15cm}
We now consider the general auction problem introduced in~\eqref{eq:market_prob}, 
Assume in round~$t$ bidder~$\ell$ is a loser with $x_\ell^*(\mathcal{B}(t))=0$. Since the allocation rule is given by an optimization problem, it can be verified that 
$$x_\ell^*(\mathcal{B}(t))=0 \implies x_\ell^*(\lbrace \mathcal{B}_{-\ell}(t), b_\ell^{k}\rbrace)=0,\ \forall k\in\mathcal{L}_t,$$
where
\begin{equation}\label{eq:gen_Lt}\begin{split}
\mathcal{L}_t=\{k\in\mathcal{K}_\ell\,\rvert\,&\X_\ell^k\subseteq\X_\ell^{k_\ell(t)},\\&\text{and}\ \forall x\in\X_\ell^k,\, b_\ell^{k}(x) \geq b_\ell^{k_\ell(t)}(x) \}. 
\end{split}
\end{equation}
In words, if $b_\ell^{k_\ell(t)}: \X_\ell^{k_\ell(t)} \rightarrow \mathbb{R}_+$ is a losing bid function, also the bid functions that lie in the epigraph of $b_\ell^{k_\ell(t)}$ are losing. Note that $\X_\ell^k\subseteq\X_\ell^{k_\ell(t)}$ is required since the bid function can be assumed to take infinitely large values outside of its domain.
The statement above holds regardless of the payment rule, and thus it is applicable to a wide-range of auctions.
\looseness=-1

In such a general framework, a generic action~$i$ is revealed either when action~$i$ is picked, or when a losing action~$j$ is picked such that action~$i$ belongs the epigraph of action~$j$. Hence, to define the revelation probabilities, we define the set of \emph{all} losing actions by $$\overline{\mathcal{L}}_t=\{k\in\mathcal{K}_\ell\,\rvert\,x_\ell^*(\lbrace \mathcal{B}_{-\ell}(t), b_\ell^{k}\rbrace)=0\},$$ and we let  $$\mathcal{R}_k=\{j\in\mathcal{K}_\ell\,\rvert\,\X_\ell^j\supseteq\X_\ell^k,\,\text{and}\,\forall x\in\X_\ell^k,\, b_\ell^j(x)\leq b_\ell^{k}(x)\},$$ be the set of all actions for which the action~$k\in\mathcal{K}_\ell$ belongs to the epigraph of the corresponding bid function. Then, according to our previous arguments, the revelation probabilities correspond to
{\medmuskip=.8mu\thinmuskip=.8mu\thickmuskip=.8mu\begin{equation}\label{eq:obsprob_general}
 \mathbf{r}_t[k] =  \begin{cases} 
 \sum_{j\in\overline{\mathcal{L}}_t\cap\mathcal{R}_k} \mathbf{w}_t[j], &\text{if }x_\ell^*(\mathcal{B}(t))=0\text{ and }k\in {\mathcal{L}}_t,\\
 \mathbf{w}_t[k], &\text{otherwise}.
 \end{cases}
\end{equation}}Observe that, in the first case, the sum involves all losing actions that can reveal information about action~$k$.

Computing these probabilities requires knowing the set of all losing actions $\overline{\mathcal{L}}_t$, which is in general unknown to the bidders.
To apply our proposed algorithm in such a general framework, we propose the following heuristic choice: 
\begin{equation*}
 \hat{\mathbf{r}}_t[k] =  \begin{cases} 
 \mathbf{w}_t[k]+\sum_{j\in \mathcal{R}_k\setminus \{\mathcal{W}_t\cup k\} } \mathbf{w}_t[j], &\substack{\text{if }x_\ell^*(\mathcal{B}(t))=0,\\ \text{and}\ k\in{\mathcal{L}}_t,}\\
 \mathbf{w}_t[k], &\text{otherwise},
 \end{cases}
\end{equation*}
where $\mathcal{W}_t\subset\mathcal{K}_\ell$ is the set of actions that lead to a non-zero allocation at every instance up to time $t$. Such set can be computed by bidder~$\ell$ based on the previous auction rounds. 
{Note that the above heuristic approximates the true $\mathbf{r}_t[k]$ by assuming that the actions that always lead to non-zero allocations ($\mathcal{W}_t$) are the ones that would lead to non-zero allocations in the current auction round. If $\mathcal{W}_t$ is equivalent to the latter set, the heuristic coincides with \eqref{eq:obsprob_general}.} 
When such revelation probabilities are plugged in \eqref{eq:our_estimator}, the resulting estimator $\tilde{\mathbf{l}}_t$ may not be unbiased and hence the result of Theorem~\ref{thm:main} is not directly applicable. Nevertheless, in our numerical case studies we will showcase the performance of the proposed heuristic in realistic electricity market models. \looseness=-1
\vspace{-.12cm}
\section{Numerical studies}\label{sec:num}
\vspace{-.12cm}
Our goal is to compare the performance of the discussed algorithms based on realistic electricity market examples. First, we consider an instance of the setting in Section~\ref{sec:simp_markt}, and show that the proposed algorithm performs significantly better than the standard bandit \textsc{Exp3}. Second, we study the IEEE 14-bus market problem from~\citep{christie2000power} to illustrate our approach discussed for the general setting. Finally, these observations are further verified by a study based on 2014 data of the two-stage Swiss reserve procurement auction from~\citep{abbaspourtorbati2015swiss}.\looseness=-1 

\vspace{-.12cm}
\subsection{Auction with Simple Constraints}\label{sec:first_ex}
\vspace{-.12cm}
Consider the set-up of Section~\ref{sec:simp_markt}, with $Q= 15$ MW and
three bidders with quadratic polynomial cost functions $c_\ell(x_\ell)=a_\ell\,x_\ell^2+d_\ell \,x_\ell$, and production limits of $\overline{X}_\ell$ MW. Values for such constants can be found in Table~\ref{tab:coeff_ex1}. Each bidder chooses among 15 actions, obtained by perturbing the linear terms of their cost functions as $d_\ell+\delta,$ where $\delta$ is chosen from a uniform distribution in the interval $[-6,\,30]$. 
Let $T=600$ be the horizon. For \textsc{Exp3} and \textsc{Hedge} algorithms, we picked the optimal learning rates $\eta= \sqrt{2\log(K) /(K T)}$ and $\eta=\sqrt{8 \log(K) / T}$, respectively, see~\citep{Bubeck2012Survey}. For our algorithm, we picked $\eta= \sqrt{2 \hat{\alpha} \log(K) /(K T)}$, by picking $\hat{\alpha}$ from $\{8.5,11,13.5\}$, since the true value of $\alpha_{\text{avg}}$ is not known a-priori {and it is required to lie in the interval $[1, K]=[1, 15]$}. These values are chosen to be the same for all bidders.
Recall that the full-information \textsc{Hedge} algorithm is unrealistic in such repeated auctions and hence we run it to upper bound the achievable performance.
\begin{table}[t!]\caption{True cost function data.}\label{tab:coeff_ex1}
  \centering 
  \resizebox{0.18\textwidth}{!}{\begin{tabular}{l c c c c}
    \toprule
    Bidder & $a_\ell$ & $d_\ell$ & $\overline{X}_\ell$  \\
    
    \midrule
    $1$ & $0.1$  & $8$ & $10$  \\
   
    $2$ & $0.095$  & $9$  & $10$  \\
   
    $3$ & $0.105$  & $10$  & $10$ \\
    \bottomrule
  \end{tabular}}
\end{table}

\begin{figure}[t!]
    \centering
    \includegraphics[width=.64\linewidth,trim={2cm 7.4cm 2cm 7.6cm},clip]{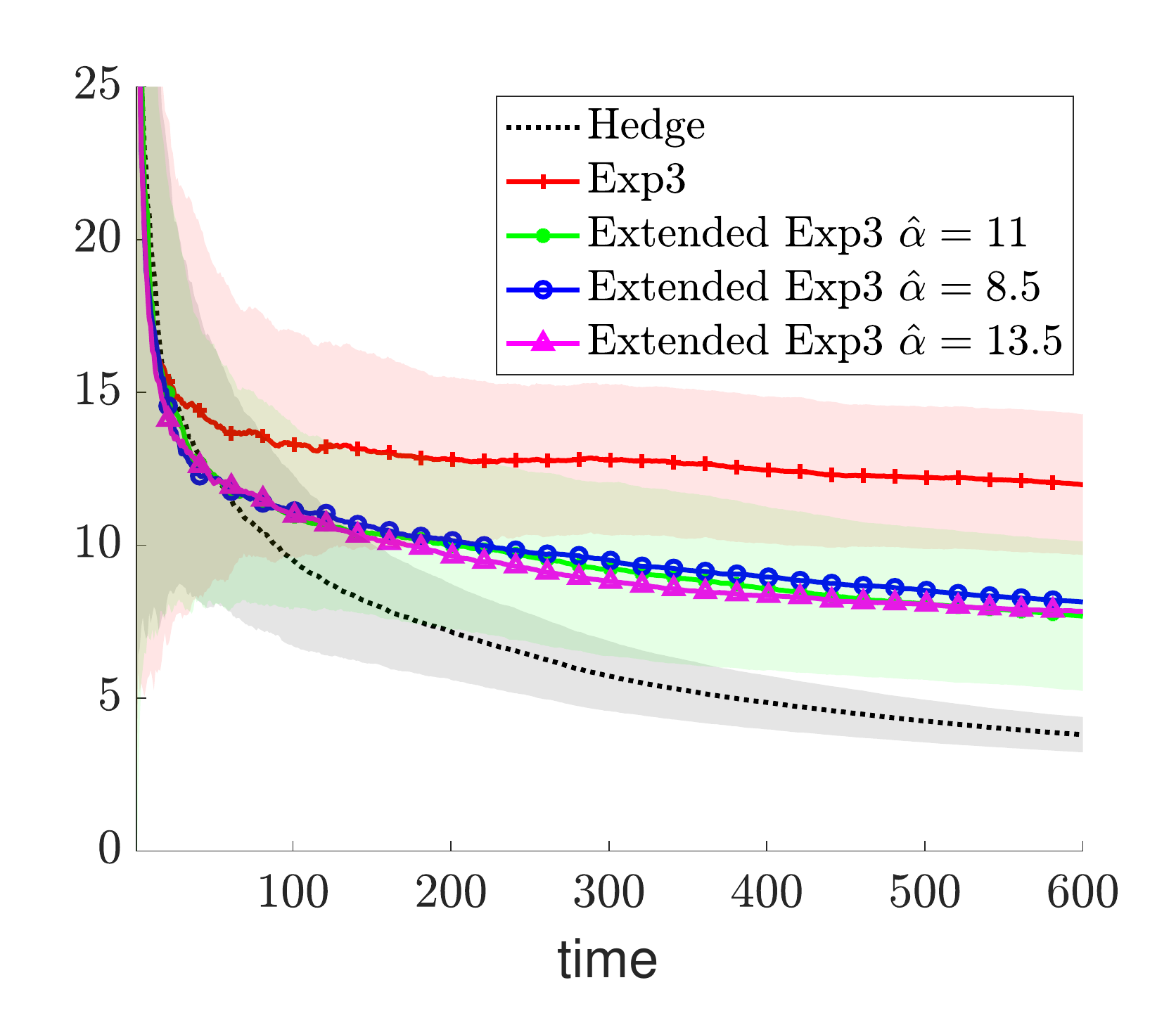}\vspace{-.1cm}
    \caption{Average regret (in \$ and averaged over 50 runs and 3 bidders) of different algorithms: Shaded areas represent $\pm$ one standard deviation.}
    \label{fig:regret_simple}
\end{figure}

\begin{figure}[t!]
    \centering \vspace{-.1cm}
    \includegraphics[width=.53\linewidth,trim={1cm 7.4cm 1cm 7.6cm},clip]{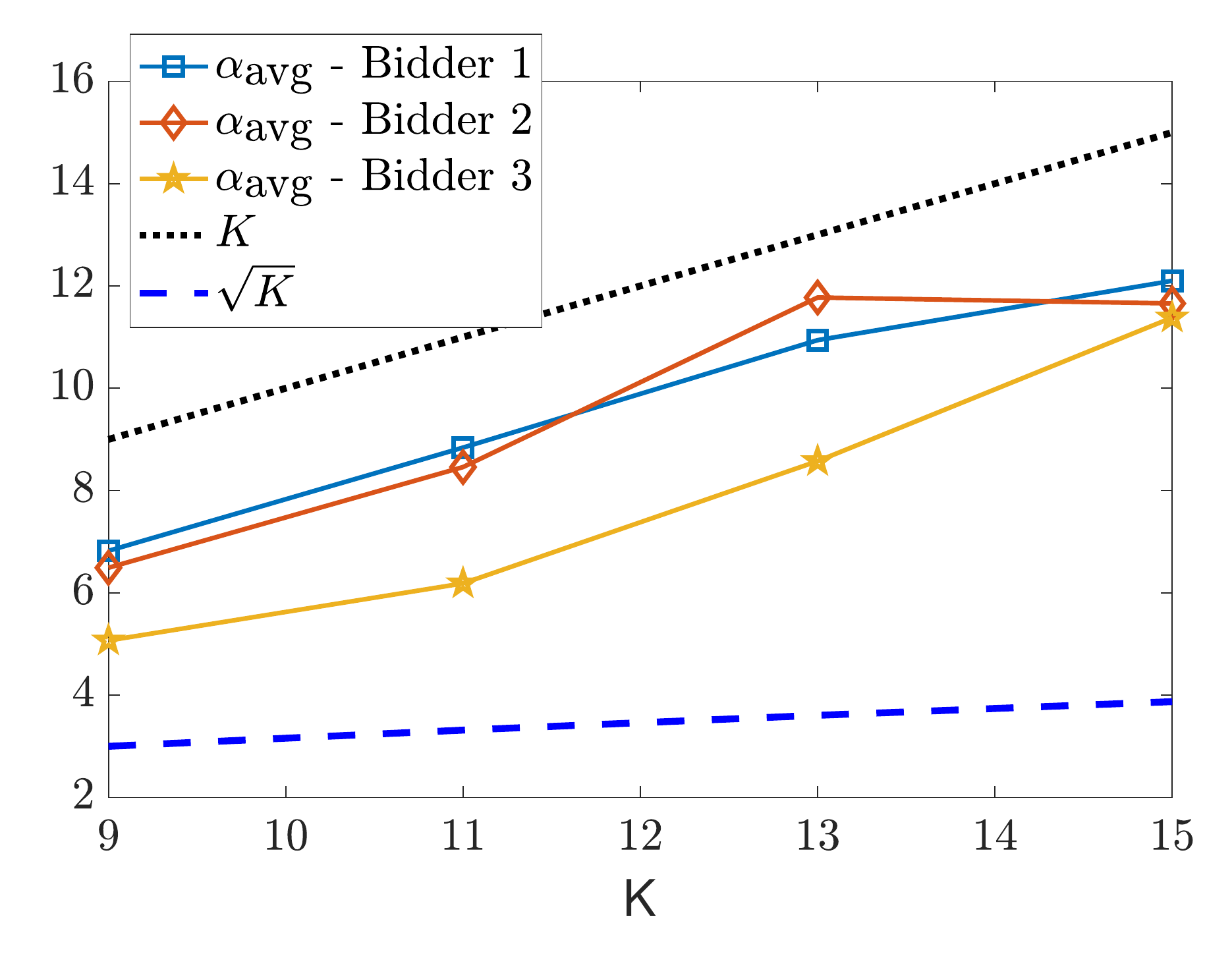}\vspace{-.1cm}
    \caption{Computed $\alpha_{\text{avg}}$ as a function of the number of actions $K$.}
    \label{fig:alpha_simple}
\end{figure}

Figure~\ref{fig:regret_simple} illustrates the average regret as in Definition~\ref{def:reg} over 50 runs and 3 bidders for the cases under which all bidders implement \textsc{Hedge}, \textsc{Exp3}, and the extended \textsc{Exp3} algorithms. Observe that the extended \textsc{Exp3} algorithm achieves a performance significantly better than \textsc{Exp3}, and also close to \textsc{Hedge} for a wide-range of learning rates $\eta$.
Figure~\ref{fig:alpha_simple} shows the average feedback information $\alpha_{\text{avg}}$ in simulations based on different number of actions, again averaged over 50 runs. We see that $\alpha_{\text{avg}}$ is close to $K$ in all cases, which explains the improvement in the bound obtained in Theorem~\ref{thm:main}. As shown in Table~\ref{tab:losing_times}, such additional information helps Bidder~2, who has the lowest production cost, to discard losing actions quickly and learn how to obtain nonzero allocations faster.

\begin{table}[t]\caption{Number of zero-allocations (averaged over 50 runs) for Bidder~2.}\label{tab:losing_times}
  \centering 
  \resizebox{0.24\textwidth}{!}{\begin{tabular}{ c c c}
    \toprule
      \textsc{Exp3} & Extended \textsc{Exp3} & \textsc{Hedge}\\
    
    \midrule
   
      $183/600$  & $131/600$ & $98/600$\\
   
    \bottomrule
  \end{tabular}}
\end{table}

\vspace{-.1cm}
\subsection{IEEE 14-Bus Optimal Power Flow Example}
\vspace{-.1cm}
The following simulations are based on the IEEE 14-bus system with DC power flow constraints, which falls in the general framework of Section~\ref{sec:gen_markt}. The true cost functions and the network model can be found in~\citep{christie2000power}. Strategy sets consists of $8$ actions which are created by perturbing the quadratic coefficient with $\delta_1$ chosen from a uniform distribution in the interval $[-0.025,\, 0.2]$, and the linear coefficient with $\delta_2$ chosen from a uniform distribution in the interval $[-10,\, 22].$ 
The VCG mechanism was proposed for such markets by~\citet{xu2015efficient,karaca2019designing}. These payments are defined as follows,\looseness=-1
\begin{equation*}
    p_\ell(\mathcal{B}(t))=b_\ell^{k_\ell(t)}(x_\ell^*(\mathcal{B}(t)))+(J(\mathcal{B}(t))-J(\mathcal{B}_{-\ell}(t))),
\end{equation*}
where $J(\mathcal{B}_{-\ell}(t))$ denotes the market cost when bidder~$\ell$ is excluded from the auction, that is, optimal value of~\eqref{eq:simple_markt} under the constraint $x_\ell=0.$

\begin{figure}[t!]
    \centering
     \includegraphics[width=.64\linewidth,trim={2cm 7.4cm 2cm 7.6cm},clip]{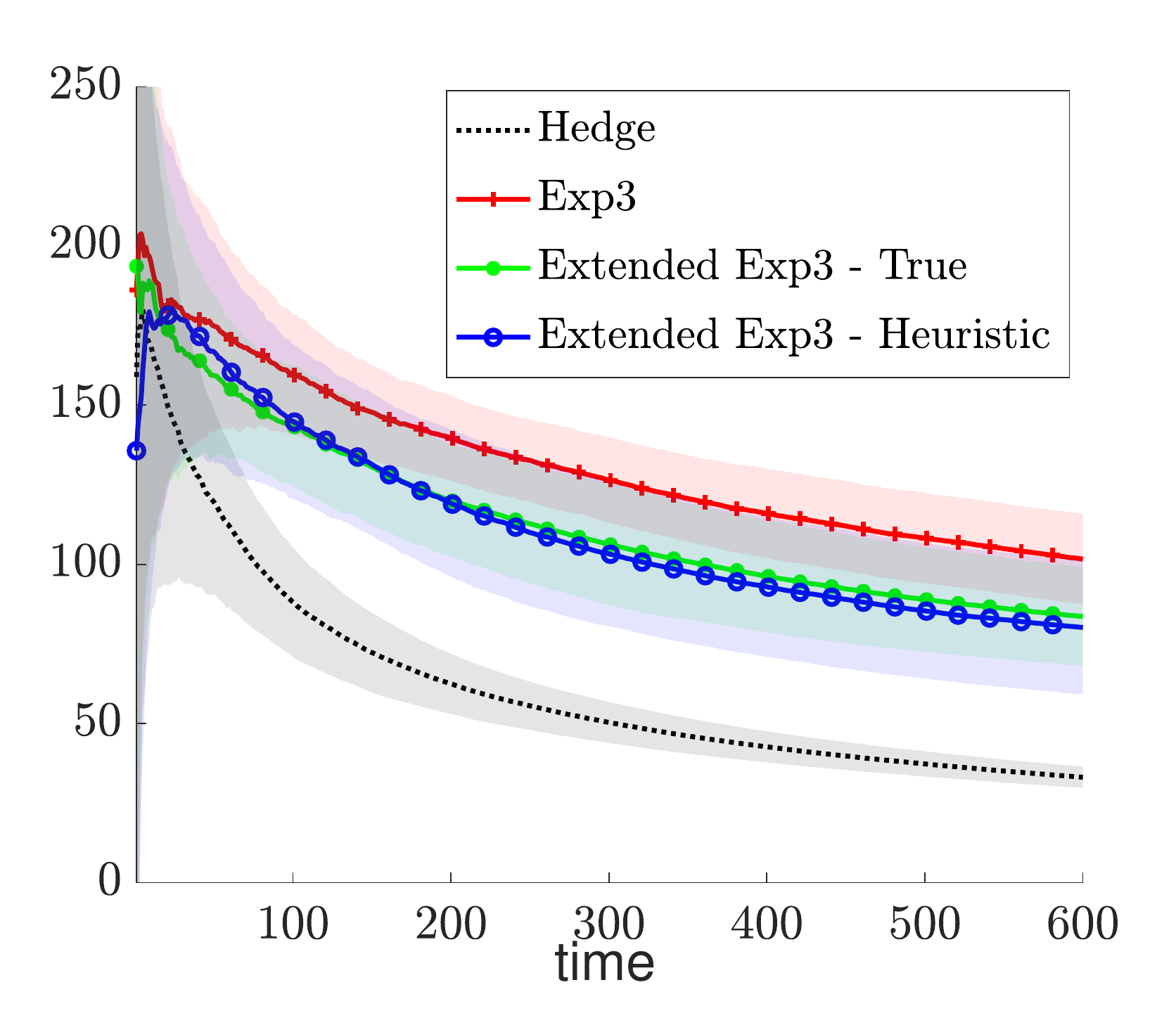}\vspace{-.1cm}
    \caption{Average regret (in \$) for IEEE 14-bus.} \vspace{-.1cm}
    \label{fig:regret_ieee}
\end{figure}

Let $T=600$ be the horizon. For \textsc{Exp3} and \textsc{Hedge} algorithms, we picked the optimal learning rates as in Section~\ref{sec:first_ex}. For our algorithm, we picked  $\eta= \sqrt{2 \hat{\alpha} \log(K) /(K T)}$ with $\hat{\alpha}=3$.
Figure~\ref{fig:regret_ieee} shows the average regrets over 50 runs for the cases under which all bidders implement \textsc{Hedge}, \textsc{Exp3}, and the extended \textsc{Exp3} algorithm. Our algorithm is implemented both using the true revelation probabilities {(by announcing $\mathcal{R}_k$ to the bidders)} and using the heuristic proposed in Section~\ref{sec:gen_markt}. The extended \textsc{Exp3} algorithm attains a better performance than the \textsc{Exp3} algorithm in both cases.  \looseness=-1
\vspace{-.1cm}
\subsection{Swiss Reserve Market}
\vspace{-.1cm}
The following simulations are based on the bids placed in the 46$^\text{th}$ weekly Swiss reserve procurement auction of 2014~\citep{abbaspourtorbati2015swiss}. This auction involves 21 plants bidding for secondary reserves, 25 for positive tertiary reserves and 21 for negative tertiary reserves. Since the problem size is big, we picked six of the largest secondary reserve providers as the learning agents.
In this auction, bids are discrete, that is, they are given by sets of reserve size and price pairs. 
Strategy sets are created by inflating the bid prices with multipliers from the set $\{1,1.05,1,15,1.25,1.35,1.45\}$. 
The market involves complex constraints arising from nonlinear cumulative distribution functions. These constraints imply that the deficit of reserves cannot occur with a probability higher than $0.2\%$.
The payment rule is  $p_\ell(\mathcal{B}(t))=b_\ell^{k_\ell(t)}(x_\ell^*(\mathcal{B}(t))).$

\begin{figure}[t!]
    \centering
     \includegraphics[width=.64\linewidth,trim={2cm 7.4cm 2cm 7.6cm},clip]{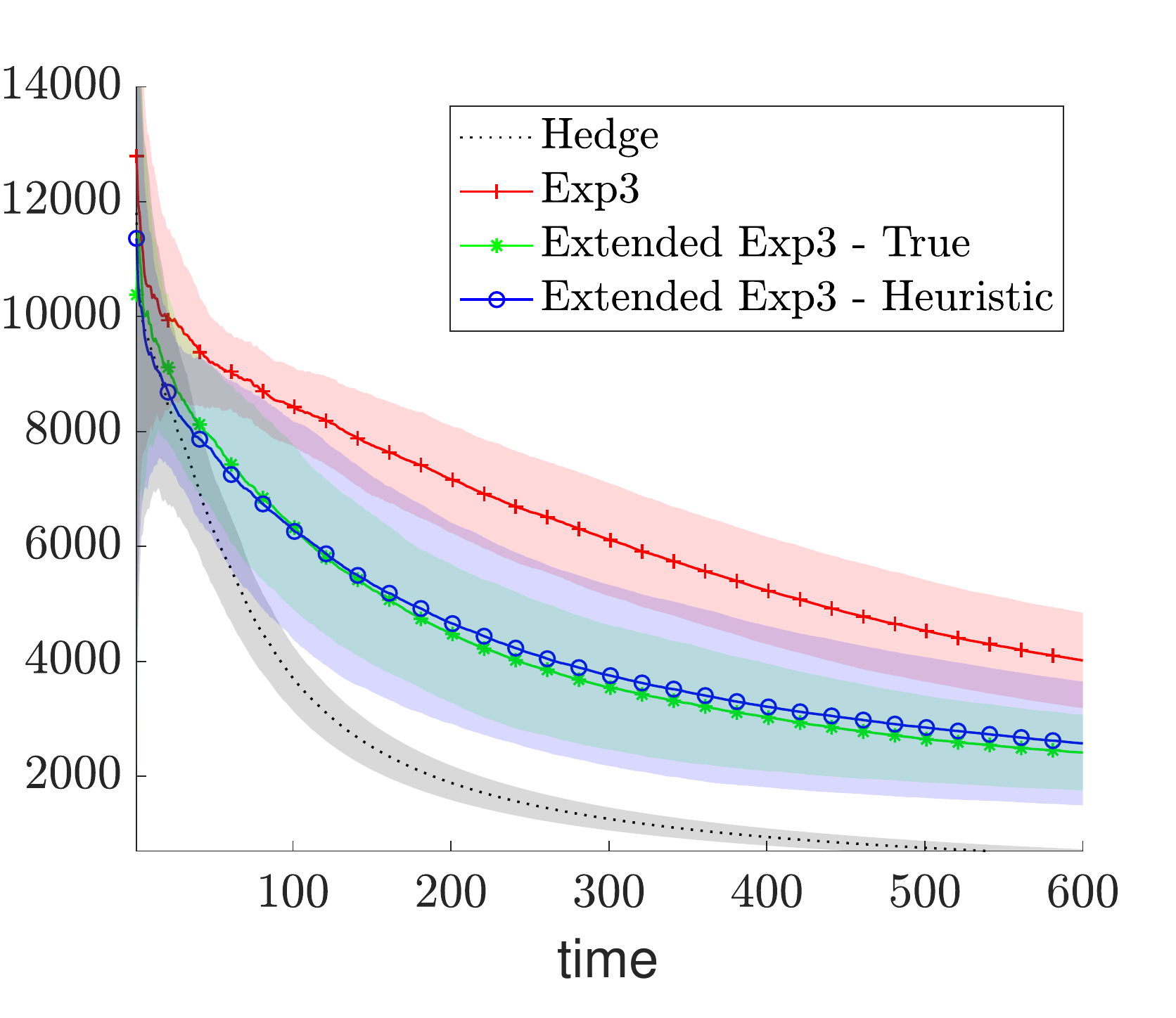}\vspace{-.1cm}
    \caption{Average regret (in CHF) for Swiss market.}
    \vspace{-.1cm}
    \label{fig:regret_swiss}
\end{figure}

Let $T=600$ be the horizon. For \textsc{Exp3} and \textsc{Hedge} algorithms, we again picked the optimal learning rates~$\eta$. For our algorithm, we picked $\eta= \sqrt{2 \hat{\alpha} \log(K) /(K T)}$ with ${\hat{\alpha}=6}$. Figure~\ref{fig:regret_swiss} shows the average regrets over 50 runs and over 6 bidders for the cases under which all bidders implement \textsc{Hedge}, \textsc{Exp3}, and the extended \textsc{Exp3} algorithm. The extended \textsc{Exp3} algorithm attains a better performance than the \textsc{Exp3} algorithm with both the true revelation probabilities {(computed by announcing $\mathcal{R}_k$)} and the heuristic choice from Section~\ref{sec:gen_markt}.
In Table~\ref{tab:ineff} we show the resulting average social cost, that is, the participants' total production cost,
averaged over $600$ auction rounds and $50$ runs.
{While there is no clear connection between participants' regret and social cost, pay-as-bid mechanism does not incentivize truthful bidding and hence no-regret algorithms increase the average social cost compared to the truthful bidding outcome.}
We will explore mechanism design with the goal of achieving lower social cost at coarse-correlated equilibria as a future work.

\looseness=-1
\begin{table}[t]\caption{{Average social cost (in CHF).}}\label{tab:ineff}
  \centering 
  \resizebox{0.4\textwidth}{!}{\begin{tabular}{c c c c}
    \toprule
      \textsc{Truthful Bidding} &\textsc{Exp3} & Extended \textsc{Exp3} & \textsc{Hedge}\\
    
    \midrule
   
     $2{,}615{,}800$  &  $2{,}652{,}300$ & $2{,}637{,}400$   & $2{,}635{,}500$\\
   
    \bottomrule
  \end{tabular}}
\end{table}

\vspace{-.12cm}
\section{Conclusion}\label{sec:conc}
\vspace{-.12cm}
In this paper, we have considered online learning in the general class of repeated auctions. For such problems, we have showed that the information structure lies in between the well-studied full-information and bandit settings. Exploiting the additional information acquired by the bidders whose bids are not accepted, we have formulated an extension to the standard bandit algorithms involving a more accurate utility estimation. We have showed that the regret guarantee of this algorithm improves upon the bandit algorithms.
Our results were verified in several case studies based on realistic electricity market models.

Our future work involves exploiting the released information in case the bids are accepted. As an extension, we are exploring the design of mechanisms with desirable properties in their coarse-correlated equilibrium.
\vspace{-.12cm}
\bibliography{ifacconf}
\vspace{-.3cm}
\appendix
\section{Proof of Theorem~\ref{thm:main}}\label{app:A}
\vspace{-.2cm}
We first bring in the following well-known second-order bound of Algorithm~\ref{alg:MWU} (MWU).
\begin{lemma}\label{lem:appx}
For any $\eta>0$ and for any sequence of loss estimates $\{\tilde{\mathbf{l}}_t\}_{t=1}^T$, we have that
\begin{equation*}
\begin{split}
    \sum_{t=1}^T \sum_{k\in\mathcal{K}_\ell} \mathbf{w}_{t}[k]\tilde{\mathbf{l}}_t[k] -& \min_{k\in\mathcal{K}_\ell}\sum_{t=1}^T \tilde{\mathbf{l}}_t[k]\\&\leq \dfrac{\log(K)}{\eta}+ \frac{\eta}{2} \sum_{t=1}^T \sum_{k\in\mathcal{K}_\ell} \mathbf{w}_{t}[k](\tilde{\mathbf{l}}_t[k])^2.\end{split}
\end{equation*}
\end{lemma}
The proof follows from standard analysis of exponentiated gradient algorithms (see \citep{cesa2006prediction, Bubeck2012Survey}). We are ready to prove our result.\vspace{-.1cm}
\begin{pf}[Proof of Theorem~\ref{thm:main}] 
First, note that the expected regret of bidder~$\ell$ can be restated as:
\begin{align*}
\mathbb{E}[R_\ell(T)] &= \mathbb{E}\left[ \sum_{t=1}^T \mathbf{l}_t[k_\ell(t)] - \min_{k\in\mathcal{K}_\ell}\sum_{t=1}^T \mathbf{l}_t[k] \right]  \\ 
& =  \sum_{t=1}^T \sum_{k=1}^K \mathbf{w}_t[k] \mathbf{l}_t[k] - \min_{k\in\mathcal{K}_\ell}\sum_{t=1}^T \mathbf{l}_t[k] \, .
\end{align*}
Moreover, since $\tilde{\mathbf{l}}_t$ is an unbiased estimate of the loss vector $\mathbf{l}_t$ (following from Fact~\ref{fact}), we have
\begin{align}
\mathbb{E}[R_\ell(T)] & = \sum_{t=1}^T \sum_{k\in\mathcal{K}_\ell} \mathbf{w}_{t}[k]\mathbb{E}[\tilde{\mathbf{l}}_t[k]] - \min_{k\in\mathcal{K}_\ell}\sum_{t=1}^T \mathbb{E}[\tilde{\mathbf{l}}_t[k]] \nonumber \\ & \hspace{-.1cm}\leq \mathbb{E} \left[\sum_{t=1}^T \sum_{k\in\mathcal{K}_\ell} \mathbf{w}_{t}[k]\tilde{\mathbf{l}}_t[k] - \min_{k\in\mathcal{K}_\ell}\sum_{t=1}^T \tilde{\mathbf{l}}_t[k] \right], \label{eq:jensen}
\end{align}
where \eqref{eq:jensen} follows from the linearity of expectation and Jensen's inequality. 
Invoking Lemma~\ref{lem:appx}, we obtain the following upper bound
\begin{equation*}
\mathbb{E}[R_\ell(T)] \leq \dfrac{\log(K)}{\eta}+ \frac{\eta}{2} \sum_{t=1}^T \sum_{k\in\mathcal{K}_\ell} \mathbf{w}_{t}[k]\mathbb{E}[(\tilde{\mathbf{l}}_t[k])^2].
\end{equation*}
Observe that, by \eqref{eq:our_estimator} and definition of $\mathbf{r}_t$, for any $k\in \mathcal{K}_\ell$  $$\mathbb{E}[(\tilde{\mathbf{l}}_t[k])^2]= \mathbf{r}_t[k] \cdot \frac{l_\ell(t)^2}{\mathbf{r}_t[k]^2} + (1-\mathbf{r}_t[k]) \cdot 0 =\frac{l_\ell(t)^2}{\mathbf{r}_t[k]}  \, .$$ This gives us
\begin{align}
\mathbb{E}[R_\ell(T)] & \leq \dfrac{\log(K)}{\eta}+\frac{\eta}{2} \sum_{t=1}^T \sum_{k\in\mathcal{K}_\ell} \mathbf{w}_{t}[k] \frac{l_\ell(t)^2}{\mathbf{r}_t[k]} \nonumber\\
& = \dfrac{\log(K)}{\eta}+\frac{\eta}{2} \sum_{t=1}^T \sum_{k\in\mathcal{K}_\ell} \frac{1}{\alpha_t^k} \label{eq:using_alphas} \\
& = \dfrac{\log(K)}{\eta}+\frac{\eta}{2} T K \frac{1}{\alpha_\text{avg}} \label{eq:using_alpha_avg}\, ,
\end{align}
where in \eqref{eq:using_alphas} we utilized the definition of $\alpha_t^k$'s and the fact that the losses are bounded in the interval $[0,1]$, and in \eqref{eq:using_alpha_avg} we utilized the definition of $\alpha_\text{avg}$.
The result of the theorem follows by substituting $\eta = \sqrt{2 \alpha_\text{avg} \log(K) /(K T)}$ in \eqref{eq:using_alpha_avg}. \QEDA

\end{pf}
\end{document}